\documentclass[amsmath,amssymb,prb,aps,showpacs,superscriptaddress,twocolumn,10pt]{revtex4-1}
\usepackage{amsmath,amsfonts,amssymb,amsthm,graphics,graphicx,epsfig,bbm}
\usepackage[colorlinks=true,citecolor=blue,linkcolor=red,urlcolor=blue]{hyperref}
\usepackage[usenames]{color}
\usepackage[english]{babel}
\usepackage{soul}
\usepackage{graphicx}
\usepackage{lipsum, babel}
\usepackage{subfigure}
\usepackage{amsmath}
\usepackage{epsfig}
\usepackage{dcolumn}
\usepackage{bm}
\usepackage{color}
\usepackage{epstopdf}
\usepackage{amssymb}
\usepackage{amstext}
\usepackage{latexsym}
\usepackage{hyperref}
\usepackage{amsfonts}
\usepackage{psfrag}
\usepackage{xcolor}
\usepackage[normalem]{ulem}
\usepackage{dsfont}
\usepackage{txfonts}
\usepackage{overpic}
\usepackage{svg}
\usepackage{lettrine}

\newcommand{\ket}[1]{\vert #1 \rangle}
\newcommand{\bra}[1]{\langle #1 \vert}

\begin{document}

\title{Quantum metamorphism}


\author{V. M. Bastidas}
\altaffiliation[]{These authors contributed equally to this work.}
\affiliation{NTT Basic Research Laboratories \& Research Center for Theoretical Quantum Physics, 3-1 Morinosato-Wakamiya, Atsugi, Kanagawa, 243-0198, Japan}
\affiliation{National Institute of Informatics, 2-1-2 Hitotsubashi, Chiyoda-ku, Tokyo 101-8430, Japan}

\author{M. P. Estarellas}
\altaffiliation[]{These authors contributed equally to this work.}
\affiliation{National Institute of Informatics, 2-1-2 Hitotsubashi, Chiyoda-ku, Tokyo 101-8430, Japan}

\author{T. Osada}
\affiliation{Tokyo University of Science, 1-3 Kagurazaka, Shinjuku, Tokyo, 162-8601, Japan}
\affiliation{National Institute of Informatics, 2-1-2 Hitotsubashi, Chiyoda-ku, Tokyo 101-8430, Japan}

\author{Kae Nemoto}
\affiliation{National Institute of Informatics, 2-1-2 Hitotsubashi, Chiyoda-ku, Tokyo 101-8430, Japan}

\author{W. J. Munro}
\affiliation{NTT Basic Research Laboratories \& Research Center for Theoretical Quantum Physics, 3-1 Morinosato-Wakamiya, Atsugi, Kanagawa, 243-0198, Japan}
\affiliation{National Institute of Informatics, 2-1-2 Hitotsubashi, Chiyoda-ku, Tokyo 101-8430, Japan}


\date{\today}

\begin{abstract}
Crystals form regular and robust structures that under extreme conditions can melt and recrystallize into different arrangements in a process that is called crystal metamorphism. While crystals exist due to the breaking of a continuous translation symmetry in space, it has recently been proposed that discrete crystalline order can also emerge in time and give raise to a novel phase of matter named discrete time crystal (DTC). In this paper, we join these two ideas and propose a model for quantum metamorphism between two DTCs of different periodicity, a $2T$ and $4T$-DTC. In our model the conditions for metamorphism come from the modulation of perturbative terms in the $4T$-DTC Hamiltonian that gradually melt its structure and transform it into a $2T$-DTC. This process is studied in detail from the viewpoint of manybody physics of periodically driven systems. 
We also propose a protocol to experimentally observe quantum metamorphism using current quantum technology.
\end{abstract}

\maketitle

\section{Introduction}

In crystallography, crystal metamorphism is generally referred as the process where a mineral becomes unstable (due to external conditions, such as pressure and temperature) and it loses its structure to give raise or recrystallize into a new crystalline arrangement \cite{Bowes1990}. Both the initial and final compounds that sit at the extremes of this process are regular atomic structures that arise from the spontaneous symmetry breaking of a discrete translation symmetry in space. While this forms the most typical idea of a crystal, that is not the only way a crystal may be formed. Unlike \emph{space}-crystals, discrete \emph{time}-crystals (or DTCs) are a novel crystalline phase of matter that break a discrete translation symmetry in the time frame (instead of space) \cite{Wilczek2012,Sacha2018,yao17,else16,Khemani2016,berdanier2018floquet,estarellas2019simulating}. The immediate consequence of this is that the dynamics of certain system's observables has a fixed (and robust) periodicity. This exotic phase of matter can only exist out of equilibrium, such as in periodically-driven systems, and it has been demonstrated in numerous experiments \cite{Zhang2017,Choi2017,Barrett2018,Smits2018,Pal2018}. In such scenarios, the system's dynamics presents a subharmonic response with respect the period of the drive that is robust to errors due to the synchronization of the manybody system's particles \cite{Zakrzewski2018,else2020discrete}.

In this context, a natural question that arises is whether phenomena typical in space-crystals, such as metamorphism, is also common in time-crystals. To address this, we will explore the quantum metamorphism in DTCs. We will specifically focus on one example, our newly proposed DTC with periodicity 4T ($4T$-DTC), with a Hamiltonian that is under the increasing modulation of perturbative terms (playing the role of the extreme conditions that allow for typical crystal metamporphism). These terms are carefully chosen such that they melt the $4T$-DTC structure to later allow for its re-crystallization into a $2T$ MBL-DTC, a discrete time-crystal of a lower symmetry (2T) that exist due to manybody localization (MBL)\cite{else16,else2020discrete}. In this work we exploit tools common in Floquet manybody physics to elucidate the insights on how the system's structure changes with increasing perturbations and slowly metamorphoses between these two DTCs. Through the evaluation of the level statistics\cite{Rigol2014,roushan17,Bastidas2018,tangpanitanon2019quantum} we observe that during the melting process there is strong level repulsion. Contrary to this, during the re-crystallization process the level repulsion disappears and there is clustering of levels. We also inspect the signatures of the metamorphosis in the fractal dimension\cite{kramer93} of the Floquet states\cite{1998GRIFONI} as well as in the Fourier spectrum of the local magnetization. Given that our model can be experimentally realized in diverse platforms (ranging from cold atoms, to superconducting qubits arrays), we end this work with an experimental proposal. Such a protocol is designed to allow for the measurement and thus observation of the metamorphic process of the DTCs as a quantum walk in configuration space, information that is accessible using \emph{state of the art} qubit devices\cite{Zhang2017,Choi2017,roushan17,RoushanChiral2017,Yan2019,Yangsen2019,chiaro2019,Martinis2019,Zha2020}.

\section{Model}

Let us begin with a simple description of our model. We consider a one-dimensional spin chain with $N$ sites governed by a time-periodic Hamiltonian of the form

\begin{figure*}[ht!]
 \centering
 \includegraphics[width=0.8\textwidth]{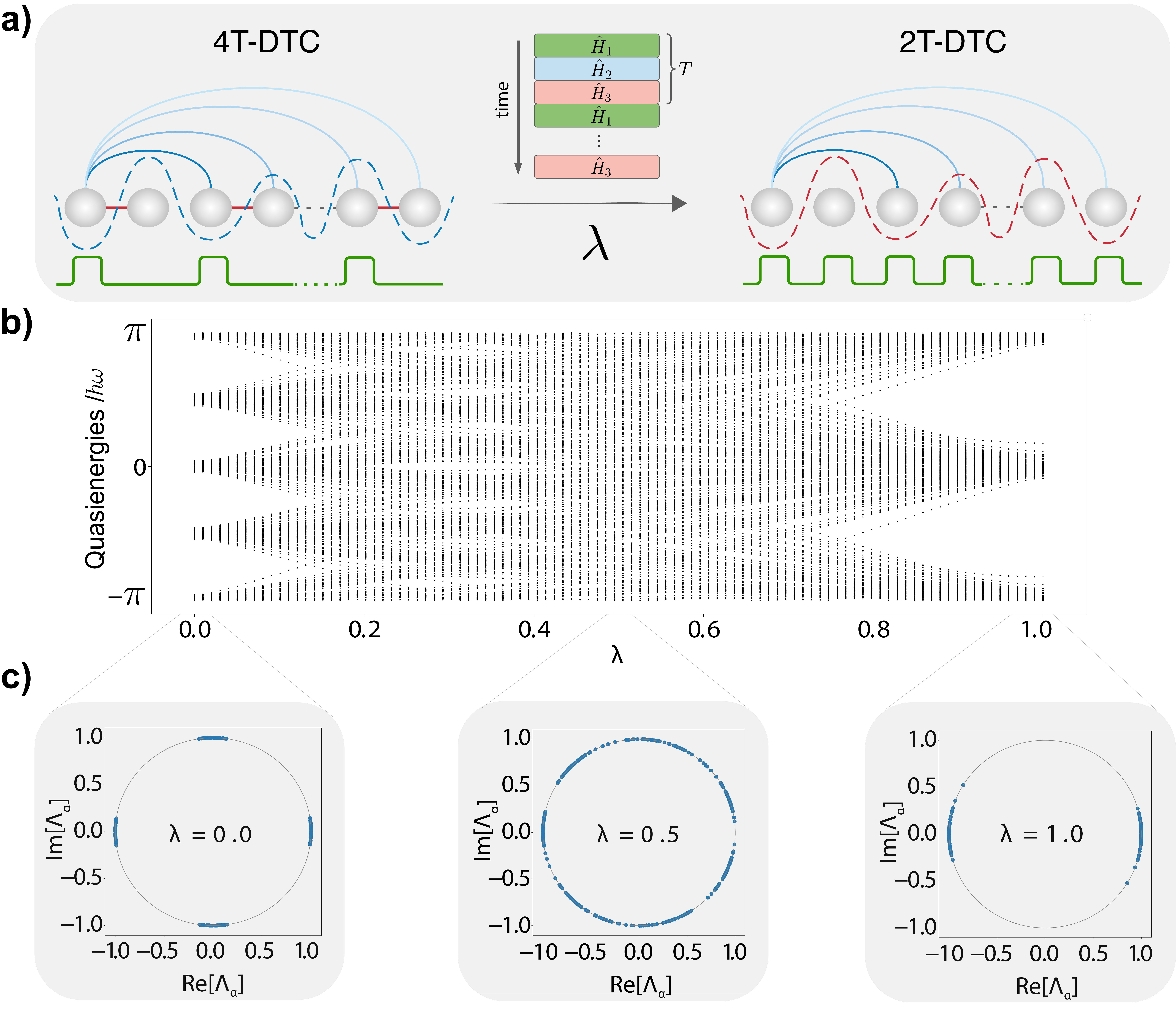}
	\caption{Spectral signatures of the discrete time crystal metamorphosis as a function of the deformation parameter $\lambda$ for a linear spin chain with $N=8$ sites. \textbf{a)} Illustration of the metamorphosis mechanism between a $4T$-DTC and $2T$-DTC. Colored elements in the spin chain represent the different actions being applied in $\hat{H}_1$, $\hat{H}_2$ and $\hat{H}_3$ for the two limiting cases of $\lambda$ (i.e. long-range and nearest-neighbor interactions, random on-site potential, and local/global $\pi$-rotations). \textbf{b)} The quasienergy spectrum as a function of the deformation parameter $\lambda$, which controls the metamorphosis between a $4T$ ($\lambda=0$) and a $2T$ ($\lambda=1$) DTCs. During the deformation, the DTC melts and recrystallizes until it is totally metamorphosed. \textbf{c)} Eigenvalues $\Lambda_{\alpha}=e^{-\mathrm{i}\varepsilon_{\alpha}T/\hbar}$ of the Floquet operator. For a $4T$-DTC, the quasienergies form four clusters ($\lambda=0.0$). During the metamorphosis, there is strong level repulsion, which leads to the melting of the discrete time crystal ($\lambda=0.5$). Subsequently, the DTC recrystallizes and the quasienergy spectrum form two clusters ($\lambda=1.0$), thus giving rise to  a $2T$-DTC.}
\label{Fig1}
\end{figure*}
 
 \begin{equation}
         \label{eq:DiscreteTimeCrystal}
  \hat{H}_{\lambda}(t)= 
  \begin{cases}
   \hat{H}_{1}= \hbar g\sum\limits^N_{l=1} \sigma_{2l-1}^{x}  +  \hbar \lambda g \sum\limits^N_{l=1} \sigma_{2l}^{x} & 0 \leq t < T_1 \\
  \hat{H}_{2}=\hbar \sum\limits^N_{l,m}J_{l,m}^{z} \sigma_{l}^{z} \sigma_{m}^{z} + \hbar (1-\lambda)\sum\limits_{l}^N W_{l}^{z} \sigma_{l}^{z} & T_1 \leq t < T_2  \\
  \hat{H}_{3}=\hbar(1-\lambda) \sum\limits_{l}^N J_{l}^{xy} (\sigma_{l}^{x} \sigma_{l+1}^{x}+\sigma_{l}^{y} \sigma_{l+1}^{y})  \\\indent\indent+
\hbar\lambda \sum\limits_{l}^N W_{l}^{z} \sigma_{l}^{z} & T_2 \leq t < T_3,

  \end{cases}
\end{equation}
whose period is $T=T_1+T_2+T_3$. Here $\sigma_{l}^\mu$ with $\mu \in \{x,y,z\}$ are the Pauli operators at the $l-$th site. The Hamiltonian $\hat{H}_1$ realizes local rotations on the two sublattices of the spin chain where the parameter $g$ is chosen such that $g T_1 = \pi/2$. The couplings $J^z_{l,m}=J_0/|l-m|^{\mu}$ decrease with the distance between two sites $l$ and $m$, with $\mu$ and $J_0$ controlling the range and strength of the interactions. In this work, we set $J_0T_2=0.15$ and $\mu=1.51$. Next $W_{l}^z \in [0,W]$ is the degree of the disorder drawn from a uniform distribution with strength $W=\pi/T_3$.  The couplings $J_{l}^{xy}$ are the strength of the flip-flop interaction between neighboring sites and they are designed in such a way that they create a dimerized chain where $J_{2l+1}^{xy}=J^{xy}$ and $J_{2l}^{xy}=0$, where $J^{xy}T_3=\pi/4$ (see \ref{Fig1}). Further the parameter $0\leq\lambda \leq 1$ acts as a control parameter for the metamorphosis between DTCs. For $\lambda=0$, the Hamiltonian \eqref{eq:DiscreteTimeCrystal} describes a $4T$-DTC. By increasing $\lambda$ the DTC melts and it can re-crystalize forming a $2T$-DTC when $\lambda=1$. For convenience we set $T_1=T_2=T_3=T/3$ through out this work.

\begin{figure*}
 \centering
 \includegraphics[width=0.99\textwidth]{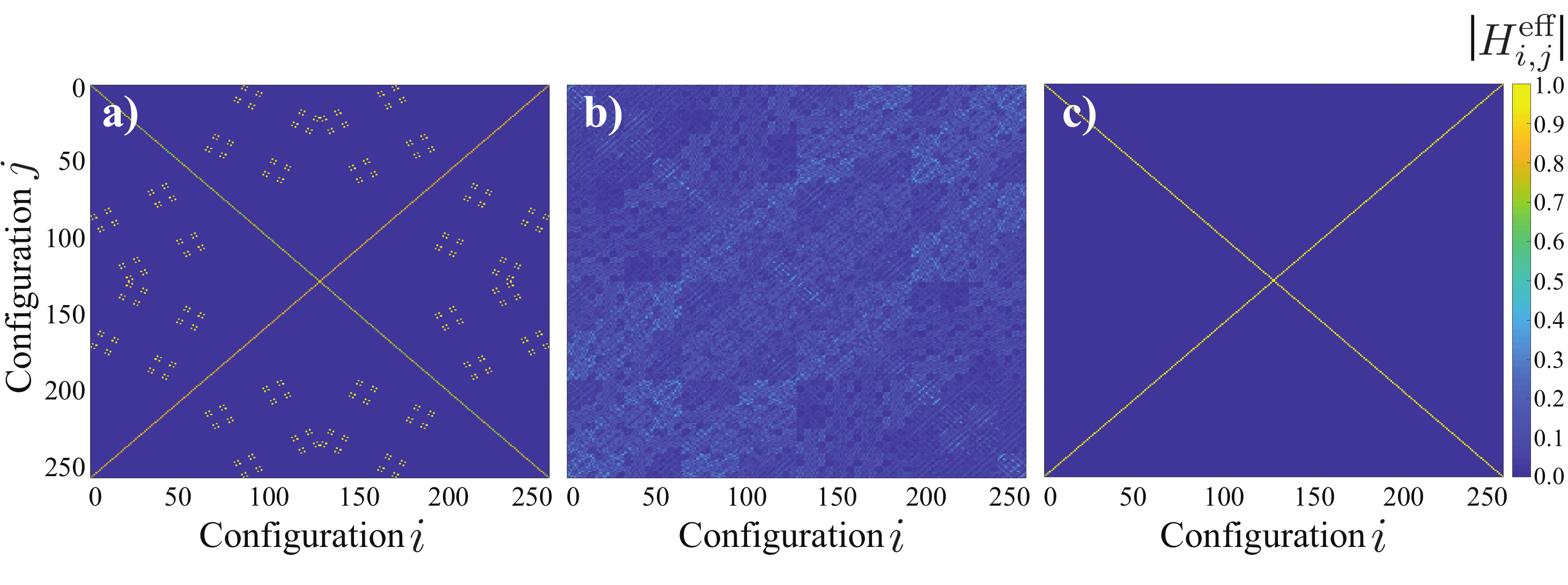}
	\caption{Entries of the effective Hamiltonian matrices for three different values of $\lambda$ for a spin chain with $N=8$ sites. \textbf{a)} For $\lambda=0$ the system is in the $4T$-DTC phase and its Hamiltonian has the form of a sparse matrix with clear regular structures. \textbf{b)} During the metamorphism the DTC melts ($\lambda=0.5$) which leads to a dense Hamiltonian matrix. \textbf{c)} If we further increase $\lambda$, the DTC recrystalizes ($\lambda=1$) giving rise to  a $2T$-DTC whose matrix has only non-zero diagonal and anti-diagonal entries.}
\label{Fig2}
\end{figure*}

\subsection{Limiting cases}
It is useful to begin by a discussion of the limiting situations of \eqref{eq:DiscreteTimeCrystal} in order to understand the qualitative features of the metamorhism of DTCs. As illustrated in Fig.~\ref{Fig1} b), in the $\lambda=0$ case, the dynamics is characterized by an initial $\pi$-rotation acting on odd sites of the lattice. After that, the system evolves under the effect of long-range interactions and disorder. At the final step of the evolution, the system is quenched and it forms dimers due to the coupling $J^{xy}_l$. If we prepare an initial fully polarized state (along the $z-axis$) $\ket{\psi(0)}=\ket{0,0,\ldots,0}$ or $\ket{\Psi(0)}=\ket{1,1,\ldots,1}$, the system will exhibit a $4T$-periodicity that appears in the longitudinal magnetization at stroboscopic times $m_l^z(nT)=\bra{\psi(nT)}\sigma^z_l\ket{\psi(nT)}$. However in the $\lambda=1$ case our Hamiltonian \eqref{eq:DiscreteTimeCrystal} describes a well known example of $2T$ MBL-DTC in the absence of error that has been experimentally realized in trapped ions \cite{Zhang2017}. For intermediate values, $\lambda$ acts as an error that melts the crystal and allows for recrystallization. Thus, the parameter $\lambda$ controls the metamorphosis between two different discrete crystalline orders in time.

\section{Spectral signatures of the metamorphosis: Level repulsion and re-crystallization of DTCs}

The Hamiltonian \eqref{eq:DiscreteTimeCrystal} is time periodic with period $T$. Due its time periodicity, it is convenient to use Floquet theory to describe the dynamics of the system~\cite{1883floquet,1998GRIFONI,Polkovnikov2015}. In Floquet theory, the dynamics at stroboscopic times $t_n=nT$ is generated by the Floquet operator~\cite{haake1991quantum,2016Restrepo} 
\begin{equation}
          \label{eq:FloquetOperator}
        \hat{\mathcal{F}}=e^{-\mathrm{i}\hat{H}_{\text{eff}}T/\hbar}=\hat{U}(T)=\hat{\mathcal{T}}\exp\left[-\frac{\mathrm{i}}{\hbar}\int_0^T\hat{H}(\tau)d\tau\right]\,
\end{equation}
where $\hat{H}(t)=\hat{H}(t+T)$, $\hat{H}_{\text{eff}}$ is referred to as the effective Hamiltonian and $\hat{\mathcal{T}}$ is the time ordering operator. The eigenvalue problem $\hat{\mathcal{F}}\ket{\Phi_{\alpha}}=e^{-\mathrm{i}\varepsilon_{\alpha}T/\hbar}\ket{\Phi_{\alpha}}$ defines the Floquet states $\ket{\Phi_{\alpha}}$ with quasienergies $-\hbar\pi/T<\varepsilon_{\alpha}<\hbar\pi/T$. Notably, the time-periodicity of Hamiltonian \eqref{eq:DiscreteTimeCrystal} is a discrete symmetry of the system~\cite{1883floquet,1998GRIFONI,Polkovnikov2015}. In the theory of discrete time crystals, some exceptional states can break the aforementioned discrete symmetry~\cite{Zakrzewski2018,else2020discrete}. In terms of the dynamics, this means that the system exhibits a subharmonic response to the external drive\cite{Sacha2018,yao17,else16,Khemani2016,berdanier2018floquet,estarellas2019simulating}. As this phenomenon occurs at stroboscopic times, it is natural to ask whether the information of a discrete time crystal appears in the quasienergies. In Fig.~\ref{Fig1}~b) we depict the quasienergies as a function of the deformation parameter $\lambda$. We clearly see that in the limiting cases discussed above, there is clustering of levels. In the case of a $4T$-DTC at $\lambda=0$, the system exhibits five clusters around quasienergies $\varepsilon_{\alpha}\approx\pm\hbar\pi/T$, $\varepsilon_{\alpha}\approx\pm\hbar\pi/2T$ and $\varepsilon_{\alpha}\approx 0$. Similarly, in the case of a $2T$-DTC  for $\lambda=1$, the system there are three clusters around quasienergies $\varepsilon_{\alpha}\approx\pm\hbar\pi/T$  and $\varepsilon_{\alpha}\approx 0$, respectively. In these extreme cases, the quasienergies are uncorrelated and the Floquet states are highly localized. Now by moving $\lambda$ away from the 0,1 extreme values, the system can explore more and more configurations of the Hilbert space. In this case the Floquet states become delocalized and there is strong level repulsion. Interestingly, Fig.~\ref{Fig1}~b) shows that for intermediate values of $\lambda$, the system can re-crystalize. Fig.~\ref{Fig1}~c) shows how the eigenvalues $\Lambda_{\alpha}=e^{-\mathrm{i}\varepsilon_{\alpha}T/\hbar}$ are grouped in 4 and 2 clusters (corresponding to the $4T$- and $2T$-DTC periodicities) for the two limiting cases $\lambda=0$ and $\lambda=1$, respectively. For the intermediate case of $\lambda=0.5$, there is strong level repulsion and no clusters are present (the DTC is completely melted).

Therefore by changing the parameter $\lambda$ we melt the crystal and allow the system to explore more configurations. The parameter $\lambda$ increases the strength of interactions in Hamiltonian \eqref{eq:DiscreteTimeCrystal} that act as defects. In this way, by subsequently breaking ergodicity, the system can recrystallize into a different structure. To have a geometrical picture of this interpretation, we have plotted the effective Hamiltonian $\hat{H}_{\lambda}^{\text{eff}}$ for three different values of $\lambda$ in  Fig.~\ref{Fig2}. Clustering of levels is accompanied by a local effective Hamiltonian as in Fig.~\ref{Fig2}~a)~and~c). For strong level repulsion, the system can explore more configurations and the effective Hamiltonian becomes non-local as seen Fig.~\ref{Fig2}~b).

\section{Localization properties of the Floquet states and their fractal dimension}
\begin{figure*}
	\centering
	\includegraphics[width=1\textwidth]{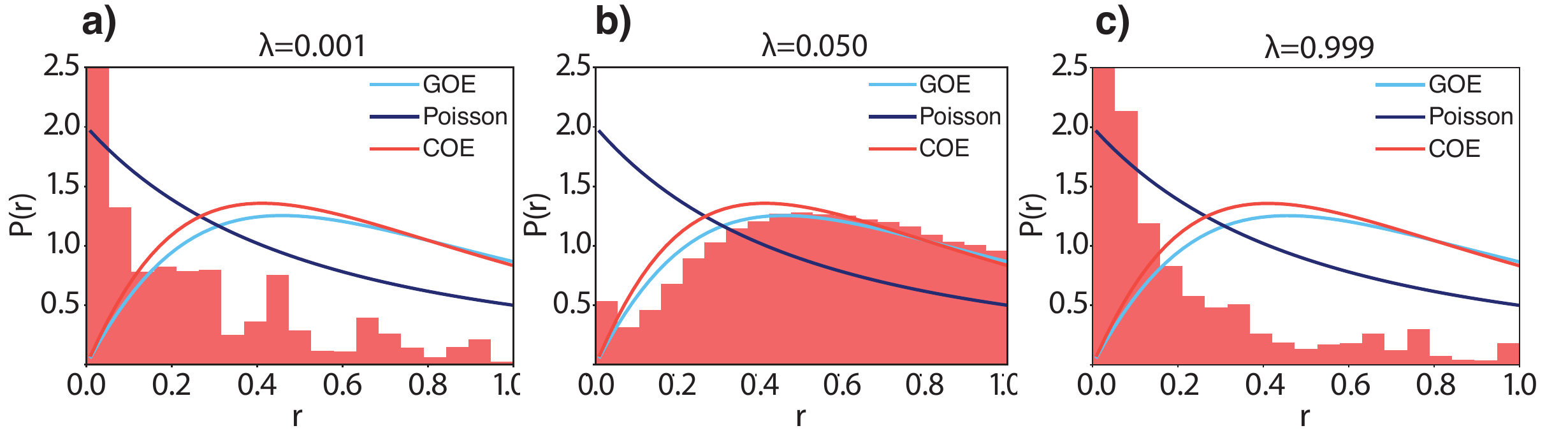}
	\caption{
		Statistical analysis of the ratios $r_{\alpha}=\min(\delta_{\alpha},\delta_{\alpha+1})/\max(\delta_{\alpha},\delta_{\alpha+1})$ between quasienergy gaps. We calculate the quasienergy spectrum of a spin chain with $N=8$ sites for $100$ realizations of disorder. a) For $\lambda=0.001$ the levels are uncorrelated and the system exhibits strong clustering of levels, which is a signature that the system is close to be integrable~\cite{haake1991quantum}. However, for $\lambda=0.001$ the level statistics is not Poissonian. b) By increasing to $\lambda=0.5$, the statistics resembles the circular orthogonal ensemble, as the system is in a melted phase. c) If we further increase to $\lambda=0.999$, the system recrystallizes to the $2T$-DTC and the statistics is closer to be Poissonian. For this value of the error there are many local conserved quantities and the system is almost integrable.
			} 
	\label{lvlstats}
\end{figure*}

\begin{figure*}
 \centering
 \includegraphics[width=0.99\textwidth]{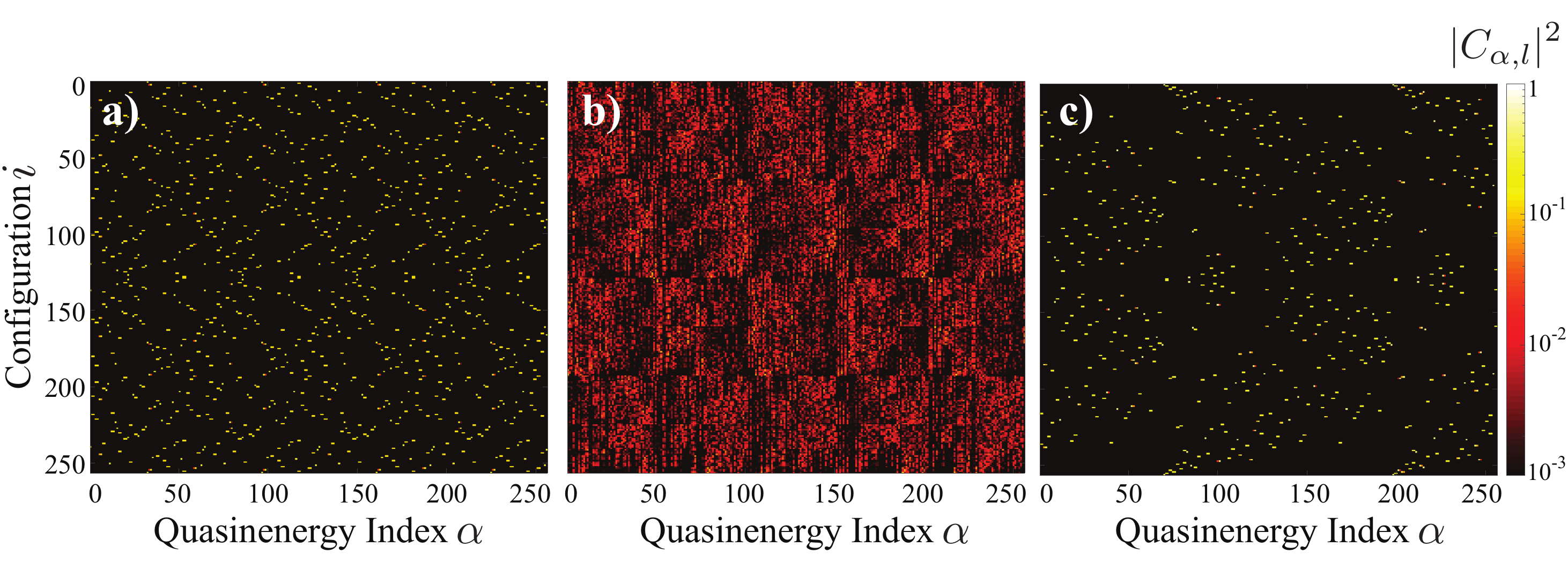}
	\caption{Localization properties of the Floquet states and metamorphism of discrete time crystals for a chain with $N=8$ sites.  The Floquet states $\ket{\Phi_{\alpha}}$ associated to the quasienergy $\varepsilon_{\alpha}$ contain valuable information about the dynamics. For this reason, here we explore their localization properties as a function of the parameter $\lambda$. To do this, for a given $\lambda$ we represent all the Floquet states as a density plot. The horizontal axis denote the index $\alpha$ associated to the quasienergies $-\hbar\omega/2\leq\varepsilon_{\alpha}\leq\hbar\omega/2$ and the vertical axis the spin configurations $l=1, \ldots, 2^N$, where $N$ is the number of spins. By using this representation one can see the populations of the different configurations for different Floquet states. \textbf{a)}  Depicts the Floquet states for $4T$-DTC with $\lambda=0$, \textbf{b)}  shows the melted DTC with $\lambda=0.5$ and  \textbf{c)}  $2T$-DTC with $\lambda=1$. This representation also provide us with a qualitative understanding of the fractal dimension. The fractal dimension of the Floquet states is related to the number of "pixels" that the wave function occupies. Notably, when the DTC is melted, the fractal dimension is high and the system explore most of the available configurations.
}
\label{Fig4}
\end{figure*}

\begin{figure}
 \centering
 \includegraphics[width=0.45\textwidth]{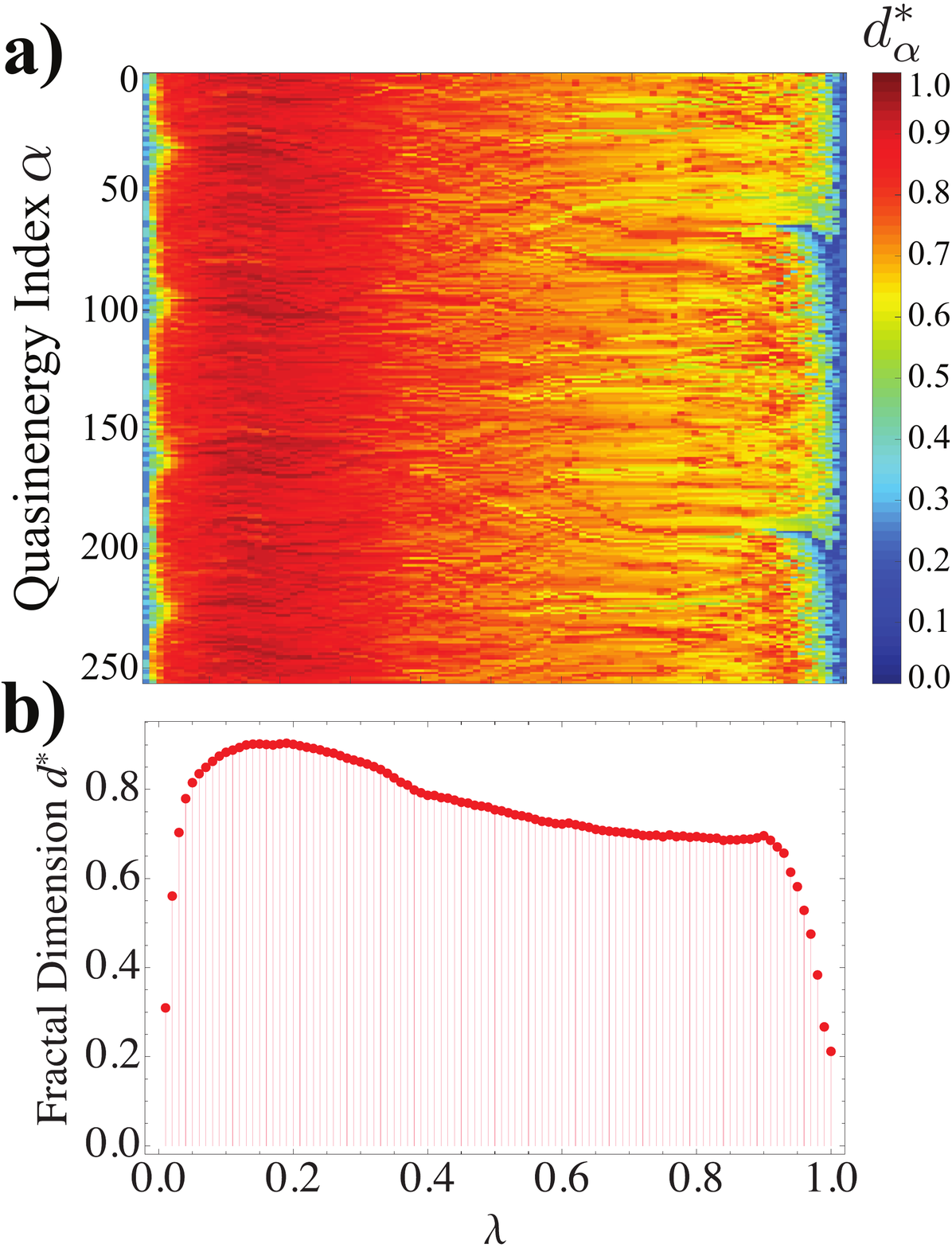}
	\caption{Fractal dimension of the Floquet states during the metamorphism of DTC's for a spin chain with $N=8$ sites. As we shown in Fig.~\ref{Fig4}, the representation of the Floquet states as a density plot give us some intuitive and qualitative understanding about how localized the Floquet states are. Now to formalize this intuition, we calculate the participation ratio $P_{\alpha}=1/\left(\sum_l |C_{\alpha,l}|^4\right)$, where $\ket{\Phi_{\alpha}}=\sum_l C_{\alpha,l}\ket{l}$ and $\ket{l}$ with  $l=1, \ldots, 2^N$ represent the different spin configurations. The fractal dimension $d_{\alpha}^*$ of a Floquet state is defined as $P_{\alpha}=D^{d_{\alpha}^*}$, where $D=2^N$ is the dimension of the Hilbert space.  \textbf{a)} Shows a density plot representing the fractal dimension $d_{\alpha}^*$ as a function of the metamorphism parameter $\lambda$ and for different Floquet states. As the fractal dimension seems quite uniform for different Floquet states, we define the average fractal dimension, as we depict in \textbf{b)}. During the metamorphosis, the fractal dimension increases when the discrete time crystal melts. When the DTC is recrystalized, the fractal dimension decreases. This is consistent with the qualitative picture discussed in Fig.~\ref{Fig4}.
	 }
\label{Fig5}
\end{figure}
 
Our Hamiltonian~(\ref{eq:DiscreteTimeCrystal}) describes an interacting system under the effect of driving and disorder. In the absence of drive, the interactions usually allow a system to explore all the configurations with a fixed energy. The disorder, on the contrary, has the ability to limit the configurations of the system that the system can reach, even under the effect of interactions. This phenomenon is referred to as manybody localization (MBL)~\cite{Altshuler2006,nandkishore15,Vosk2015,khemani17,abanin17}. However, the situation is dramatically changed when the system is under the effect of an external drive~\cite{Abanin2015,Bastidas2018,Zha2020}. That is, the conservation of energy is broken and the system now can explore a bigger set of configurations than in the undriven case~\cite{Rigol2014,Saito2016}.

\subsection{Localization and Statistical properties of the quasienergy spectrum}
When dealing with disordered manybody systems one typically encounters a competition between interactions and disorder~\cite{Altshuler2006,abanin17,Huse2014,roushan17}. Interactions distribute the energy between all the available states, because they usually break conserved quantities at the single-particle level, leading to thermalization and ergodic behavior~\cite{Srednicki1994,dalessio16}. In such scenarios, the destruction of conserved quantities induces strong correlations between the energy eigenvalues giving raise to a strong level repulsion~\cite{haake1991quantum,dalessio16,roushan17}. On the other hand, disorder can suppress ergodicity leading to a manybody localized phase, which is characterized by an extensive number of conserved quantities~\cite{nandkishore15,Vosk2015,khemani17} and the presence of uncorrelated energy levels~\cite{haake1991quantum,dalessio16,roushan17}. In order to explore these behaviors during our metamorphism process we make use of level statistics, a well-established diagnosis for thermalization and manybody localization in condensed-matter physics~\cite{Abanin2015,abanin17,Rigol2014,roushan17,Bastidas2018}.

In the absence of an external drive, one is interested in the repulsion between neighboring energy levels. To do this, one usually needs to perform spectral unfolding in order to study fluctuations about the average integrated density of levels~\cite{haake1991quantum}. This procedure can be cumbersome, because it requires some intuition of the smooth part of the density of levels of the system~\cite{haake1991quantum}. Recently, it has been shown that the unfolding of energy levels is not required and instead, we can use the distribution of ratios $P(r)$ with $r_{s}=\min(\delta_{s},\delta_{s+1})/\max(\delta_{s},\delta_{s+1}) \leq 1.0$ with $\delta_{s}=E_{s+1}-E_{s}$, where $E_{s}$ are the energy eigenvalues sorted in order of increasing value~\cite{dalessio16,Rigol2014}. In the MBL phase, the levels are uncorrelated because the manybody states are highly localized in space and the level statistics follows a Poissonian-like level statistics~\cite{Rigol2014,tangpanitanon2019quantum}
\begin{equation}
\label{eq:PoissonDist}
P_{\text{Poisson}}(r)=\frac{2}{(1+r)^2}
\ .
\end{equation}
When the interactions overcome the effect of disorder, the aforementioned conserved quantities are destroyed and the system is able to explore all the configuration space with a constant energy~\cite{dalessio16}. This is reflected in a strong level repulsion because some states are delocalized. As a consequence, in the case of real Hamiltonians, the system follows a universal level statistics
\begin{equation}
\label{eq:GOEDist}
P_{\text{GOE}}(r)=\frac{27}{4}\frac{r+r^2}{(1+r+r^2)^{5/2}}
\end{equation}
associated with the Gaussian Orthogonal ensemble (GOE) of random matrices~\cite{dalessio16,roushan17}.

In this work we are interested in the melting of a $4T$-DTCs and its recrystallization into a $2T$-DTC under the effect of a parameter $\lambda$ as shown in the Hamiltonian from Eq.~\ref{eq:DiscreteTimeCrystal}. Below we show that the time-crystalline order is directly related to the presence of an MBL phase. In the absence of error, the time crystal has several conserved quantities and the motion is periodic. However, discrete time crystals are quantum phases of matter that appear in periodically-driven quantum systems and energy is not conserved anymore. Therefore,  we have to work with gaps $\delta_{\alpha}=\varepsilon_{\alpha+1}-\varepsilon_{\alpha}$, where $-\hbar\pi/T<\varepsilon_{\alpha}<\hbar\pi/T$ are the quasienergies and $T$ is the period of the drive. Under the effect of a small error from varying the $\lambda$ parameter, there is a small coupling between different symmetry multiplets and the quasienergy level statistics follows a Poissonian behavior as in Eq.~\eqref{eq:PoissonDist}. As a consequence, the system is in the MBL phase, which protects the system from heating up to infinite temperatures~\cite{Rigol2014,Khemani2016}. When $\lambda$ is increased, the conserved quantities are destroyed and the Floquet states become highly delocalized in the configuration space. The DTC melts and an MBL-to-ergodic phase transition takes place. The statistics of levels in this case is given by

\begin{eqnarray}
\label{eq:COEDist}
P_{\text{COE}}(r)=\frac{2}{3}\left\{\left[\frac{\sin\left(\frac{2\pi r}{r+1}\right)}{2\pi r^2}\right]+\frac{1}{(1+r)^2}+\left[\frac{\sin\left(\frac{2\pi}{r+1}\right)}{2\pi }\right]\right\}\\
\nonumber-\frac{2}{3}\left\{\left[\frac{\cos\left(\frac{2\pi }{r+1}\right)}{2\pi r^2}\right]+\left[\frac{\cos\left(\frac{2\pi r}{r+1}\right)}{r(r+1)}\right]\right\}
\,
\end{eqnarray}
which has the same statistics as the circular orthogonal ensemble of random matrices~\cite{Rigol2014,tangpanitanon2019quantum}. 

In quantum systems, quantum signatures of chaotic dynamics lead to a universal behavior of the level statistics giving rise to distributions $P_{\text{GOE}}(r)$ and $ P_{\text{COE}}(r)$ for undriven and driven systems~\cite{Rigol2014,tangpanitanon2019quantum}, respectively. 
In the context of time crystals, when we increase $\lambda$, we see a crossover between $P_{\text{Poisson}}(r)$ and $P_{\text{COE}}(r)$, because $\lambda$ destroys conserved quantities. As we increase $\lambda$ further, we observe that the level statistics is closer to $P_{\text{COE}}(r)$.  Fig.~\ref{lvlstats} depicts the results for the level statistics of the melting of a $4T$-DTC and its recrystallization to a $2T$-DTC by modulation of a parameter $\lambda$ for $\lambda=0.001$, $\lambda=0.5$ and $\lambda=0.999$. Our results therefore show that at the extreme $\lambda$ points, the two time-crystalline phases, are protected by MBL.
\subsection{Localization of Floquet states and the fractal dimension}
 After discussing in detail spectral signatures of localization, in this section we focus on a well known diagnosis of localization of Floquet states known as the fractal dimension~\cite{kramer93}. Let us begin by considering a Floquet state $\ket{\Phi_{\alpha}}=\sum_l C_{\alpha,l}\ket{l}$, where $\ket{l}$ is a basis composed of $2^N$ spin configurations of the system. To be able to quantify the degree of localization of a state in a given basis we use the participation ratio~\cite{kramer93,roushan17,Bastidas2018}, a fundamental concept in the theory of Anderson localization~\cite{anderson58}. Given a Floquet state $\ket{\Phi_{\alpha}}$, the participation ratio is defined as $P_{\alpha}=1/\sum |C_{\alpha,l}|^4$. When a Floquet state is fully localized in a given configuration $\ket{l_0}$, then $C_{\alpha,l_0}=1$ and the participation ratio is $P_{\alpha}=1$. This means that only one configuration participates in the dynamics. On the contrary, fully delocalized states behave like plane waves in the configuration space with $C_{\alpha,l}=2^{-D/2}$, where $D=2^N$ is the dimension of the Hilbert space. Thus, the participation ratio is $P_{\alpha}=D$. In the theory of Anderson localization~\cite{anderson58,kramer93}, it is common practice to define the fractal dimension $d_{\alpha}^*$ as $P_{\alpha}=D^{d_{\alpha}^*}$. In these terms, localized states are zero dimensional objects and delocalized ones have fractal dimension $d_{\alpha}^*=1$. 
 
Next let us briefly discuss the intimate relation between the fractal dimension and level statistics, which is of utmost importance to understand the metamorphism of discrete time crystals. When a driven system is in the fully-localized MBL regime, it exhibits an extensive set of local integrals of motion~\cite{abanin17,Abanin2015}. Thus, the Floquet states are localized and the corresponding quasienergies are uncorrelated, which gives rise to a Poissonian level statistics~\cite{Rigol2014,tangpanitanon2019quantum,Bastidas2018}. In turn, we also expect the fractal dimension to be small, because due to the contraints imposed by the conserved quantities, ergodicity is broken and the system cannot explore the whole Hilbert space. On the contrary, in the ergodic regime, the drive breaks the conserved quantities and there is a strong level repulsion of quasienergies, whose level statistics follow a COE distribution~\cite{Rigol2014,tangpanitanon2019quantum,Bastidas2018}. As the conserved quantities are broken, the system can explore more configurations and the fractal dimension should increase.

To have a geometrical representation of the Floquet states during the metamorphosis process, we depict the probabilities $|C_{\alpha,l}|^2$ as a density plot, where the axis are the configuration index $l$ and the quasienergies $-\hbar\pi/T<\varepsilon_{\alpha}<\hbar\pi/T$. Fig.~\ref{Fig4} illustrate the results for the metamorphism with $\lambda=0.001$ (4T-DTC), $\lambda=0.5$ (melted DTC) and $\lambda=0.999$ (2T-DTC). This provides us with a simple geometrical picture of the localization in terms of the pixels of the images. Intuitively, when a system is in the DTC phase, few configurations participate in the dynamics, in contrast to a melted DTC. To formalize this intuitive picture, we numerically calculate the fractal dimension $d_{\alpha}^*$ and plot it as a function of $\lambda$ and the quasienergy index $\alpha$ in Fig.~\ref{Fig5}~a). Our numerical result confirms our intuition. For small values of the metamorphism parameter $\lambda$, the fractal dimension is small. As we increase $\lambda$, the fractal dimension increases when the DTC melts. Correspondingly, when the system recrystalizes into a 2T-DTC, the fractal dimension decreases due to the emergent conserved quantities associated to the MBL DTC. Fig.~\ref{Fig5}~b), we depict the average of the fractal dimension $d^*=1/D\sum_{\alpha}d_{\alpha}^*$ as a function of $\lambda$.
 
\section{Signatures of metamorphism of DTCs in the dynamics of the local magnetization}

\begin{figure*}[ht!]
 \centering
 \includegraphics[width=0.7\textwidth]{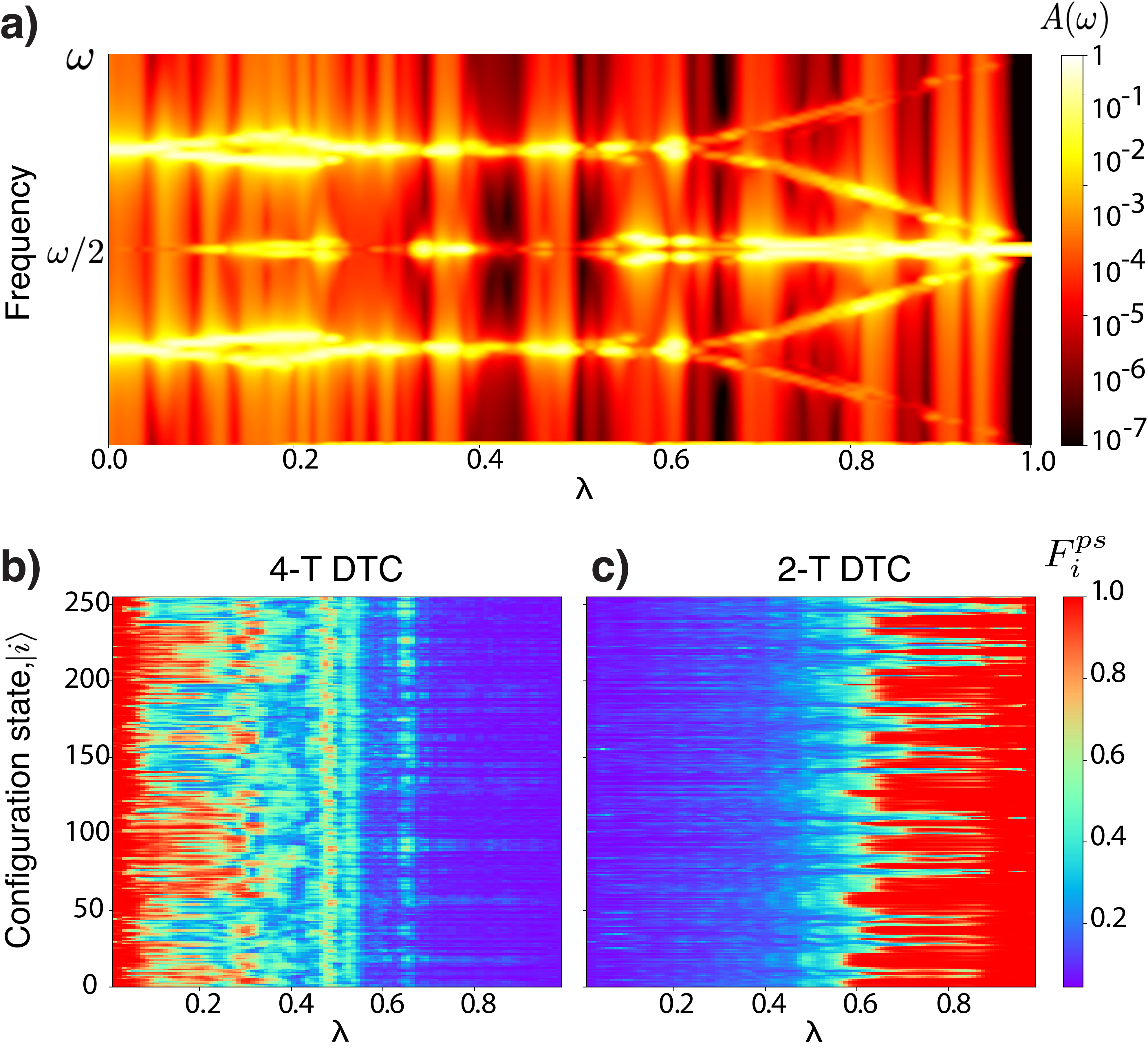}
	\caption{Fourier spectrum and fidelities of the power spectrum during the metamorphosis of an 8 site spin chain under the effect of a driving with frequency $\omega=2\pi/T$. a) Plot of the Fourier spectrum $A(\omega_k)=|\mathcal{M}_{i,Z}{k}|^2$ for different values of the parameter $\lambda$. For small values of $\lambda$, the peaks at frequencies $\omega_k=\omega/4$ and $\omega_k=3\omega/4$,  we can see signatures of the 4T-DTC. By increasing $\lambda$, the DTC melts and there are more frequencies in the Fourier spectrum. For $\lambda$ closer to $1$, the DTC recrystalized, which can be seen in the peak at $\omega_k=\omega/2$. b) and c) Depict the fidelities $F_{i}^{\text{ps-4T}}(\lambda)$ and $F_{i}^{\text{ps-2T}}(\lambda)$ of the Fourier spectrum for $4T-$ and $2T-$DTCs, respectively. We show the fidelities as a function of $\lambda$ and for different configurations $\ket{i}$, labelled by $i=1,\ldots,2^N$.
	These plots clearly show that some configurations are less robust for different values of $\lambda$. 
	}
\label{Fig6}
\end{figure*}

In order to shed some light into this new mechanism of time crystals' metamorphism we explore the power spectrum (or Fourier spectrum) as a tool to characterize the subharmonic response of the system. For these, we first obtain the expectation value of the total magnetization $M_{i,Z}(mT)=1/N\sum_{r=1}^N\langle\psi(mT)|\sigma_r^z|\psi(mT)\rangle$ at stroboscopic times, where $|\psi(0)\rangle=|i\rangle$. For a given initial configuration $i$, one can record the measurements $M_{i,Z}(mT)$, thus obtaining a time series 
 \begin{equation}
 \label{eq:TimeSeries}
 \{M_{i,Z}(0),M_{i,Z}(T),\dots,M_{i,Z}(nT)\}
 \ ,
 \end{equation}
where $nT$ is the number of periods that the system evolve. With the data for the time series at hand, we can calculate the discrete Fourier transformation
\begin{equation}
 \label{eq:FourierTransverseMagPopulation}
\mathcal{M}_{i,Z}(k)=\frac{1}{n}\sum_{m=1}^ne^{-\frac{\mathrm{i}2\pi k}{n}m}M_{i,Z}(mT)=\frac{1}{n}\sum_{m=1}^n e^{-\mathrm{i}\omega_k m T}M_{i,Z}(mT)
\ ,
\end{equation}
where $\omega_k=2\pi k/nT$ and  $k\in[0,n-1]$. For a given coefficient $\lambda$, the associated power spectrum 
\begin{equation}
 \label{eq:FourierPowerSpec}
\boldsymbol{V}_{i,\lambda}=\{|\mathcal{M}_{i,Z}(0)|^2,|\mathcal{M}_{i,Z}(1)|^2\dots,|\mathcal{M}_{i,Z}(n-1)|^2\}
\end{equation}
indicates how strong is the contribution of a the $k-$th harmonic to the time series. Note that for convenience, we have arranged the Fourier coefficients in the form of a vector $\boldsymbol{V}_{i,\lambda}$. In the absence of error, all the configurations show a magnetization that varies as $M_{i,Z}(mT)=(-1)^mM_{i,Z}(0)=e^{\mathrm{i}m\pi}M_{i,Z}(0)
$ and the power spectrum shows two peaks at frequencies $\omega=n/4$ and $\omega=3n/4$ for the $4T$-DTC and a single peak at a frequency $\omega_{n/2}$ for the $2T$-DTC. These peaks, shown in Fig.~\ref{Fig6} a), are clear signatures of the subharmonic response of the manybody system. As we discussed, the periodic motion is related to local integrals of motion~\cite{serbyn13,Zha2020}. The DTC is protected against heating due to MBL~\cite{else2020discrete}. However, as we increase the parameter $\lambda$, some configurations are more affected than others and they loose their periodic nature~\cite{estarellas2019simulating}. To have a quantitative way of looking at this, we calculate the fidelities $F_i^{\text{ps-4T}}(\lambda)$ and $F_i^{\text{ps-2T}}(\lambda)$ of the power spectrum~\cite{estarellas2019simulating}, defined as follows
\begin{equation}
 \label{eq:Fidelity1}
F_{i}^{\text{ps-4T}}(\lambda)=\sqrt{\frac{\boldsymbol{V}_{i,0}\cdot\boldsymbol{V}_{i,\lambda}}{\|\boldsymbol{V}_{i,0}\|\:\|\boldsymbol{V}_{i,\lambda}\|}}
\ .
\end{equation}
\begin{equation}
 \label{eq:Fidelity2}
F_{i}^{\text{ps-2T}}(\lambda)=\sqrt{\frac{\boldsymbol{V}_{i,1}\cdot\boldsymbol{V}_{i,\lambda}}{\|\boldsymbol{V}_{i,1}\|\:\|\boldsymbol{V}_{i,\lambda}\|}}
\ .
\end{equation}
These very simple quantities tell us how far the power spectrum of the system at an arbitrary value of $\lambda$ is from the ideal $4T$-DTC ($\lambda=0$) and $2T$-DTC ($\lambda=1$). In Fig.~ \ref{Fig6} \textbf{b)} and \textbf{c)} we present these two fidelities as a function of the configuration state $\vert i\rangle$ and the parameter $\lambda$. This shows clear evidence that not all the configurations are stable under the effect of a rotation error.


\begin{figure*}[ht!]
 \centering
 \includegraphics[width=0.99\textwidth]{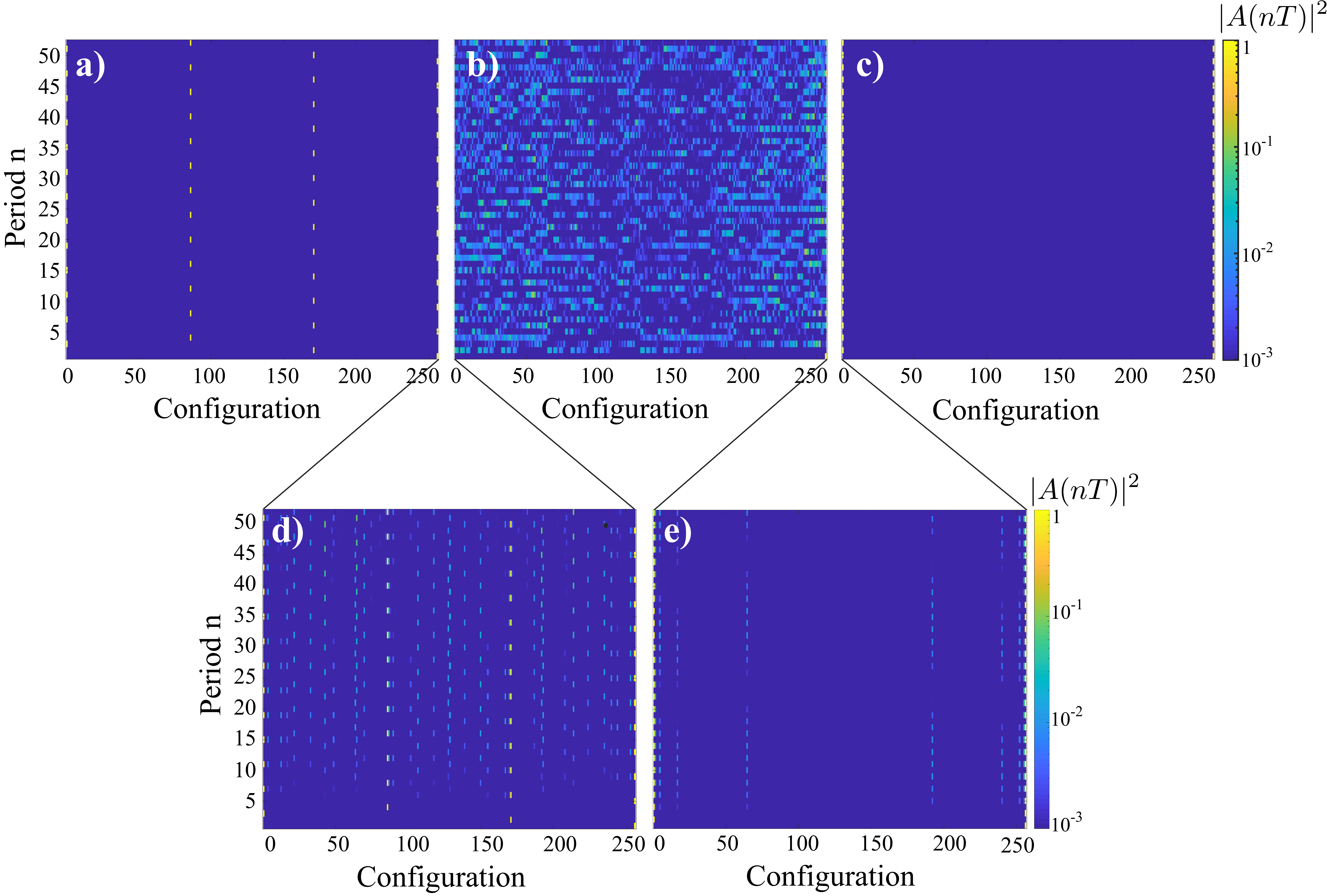}
	\caption{Dynamics of the populations $|A(nT)|^2$ as a quantum walk in the configuration space as an experimental protocol to measure the metamorphosis between a $4T$-DTC and $2T-DTC$ of $N=8$ sites. a) For $\lambda=0$, the system explores just four configurations of the Hilbert space. b) With intermediate values of $\lambda$, the time-crystal melts and becomes highly ergodic, allowing the system to visit many of the possible configuration states. c) Once recrystallized when $\lambda=1$, the system localizes and its dynamics explores only two configurations. Panels d) and e) show the quantum walk when the system slightly deviates for the two limiting cases ($\lambda=0.01$ and $\lambda=0.99$, respectively).}
\label{Fig7}
\end{figure*}
 
 \section{Quantum walks in the configuration space}
In the previous sections we have discussed in detail how the parameter $\lambda$ acts as an error that breaks the conserved quantities and allows the system to recrystallize into another DTC during the metamorphism. We have also shown that the metamorphism can be experimentally observed by measuring the longitudinal magnetization at stroboscopic times $m_l^z(nT)=\bra{\psi(nT)}\sigma^z_l\ket{\psi(nT)}$ when we prepare a $z-axis$ fully polarized initial state $\ket{\psi(0)}=\ket{0,0,\ldots,0}$ or $\ket{\psi(0)}=\ket{1,1,\ldots,1}$. In the limiting cases of $\lambda=0$ and $\lambda=1$, the magnetization is periodic with periods $4T$ and $2T$, respectively. For these values of $\lambda$, the dynamics of the system takes place within a restricted subspace of the Hilbert space, because the transitions to other configurations are suppressed by conservation rules.

By increasing $\lambda$, the driving induces resonances between multiple configurations, which cannot be captured by the observable $m_l^z(nT)$. To overcome this limitation, we use a method described in the supplemental material of Ref.~\cite{estarellas2019simulating} which discussed how to experimentally access properties of the time crystal in terms of quantum walks at stroboscopic times in a configuration space with $2^N$. 

The idea behind the quantum walk in configuration space is quite simple. At stroboscopic times $t_n=nT$, the dynamics of the system are governed by the effective Hamiltonian
\begin{equation}
      \label{eq:EffHamiltonianConf}
     \hat{H}^{\text{eff}}=\sum_{l}E_l\ket{l}\bra{l}+\sum_{l,m}V_{l,m}\ket{l}\bra{m}
     \ ,
\end{equation}
where $E_l$ denotes the effective local energies associated to the $l$-th configuration and $V_{l,m}$ denotes the couplings between different configurations. When the energy detuning $|E_l-E_m|$ between two configurations is smaller the coupling $V_{l,m}$, there is a resonance and the configurations become hybridized~\cite{Roy2019,estarellas2019simulating}. This representation is very useful, because now we can visualize the dynamics of the system as the motion of a particle in a lattice with $D=2^N$ sites. Thus, the system is able to tunnel to resonant configurations and perform a quantum walk at discrete times.

During the metamorphism of DTCs, we vary a parameter $\lambda$ that acts as an error. For small values of $\lambda$, there are several conserved quantities and the effective Hamiltonian~\eqref{eq:EffHamiltonianConf} is local in the configuration space, as depicted in Fig.~\ref{Fig2}~a). Thus, there are few resonant configurations and the dynamics takes place mostly between 4 configurations, which is related to the period of the DTC. By increasing $\lambda$, the drive induces resonances and the system can explore more configurations, which leads to a non-local effective Hamiltonian [see Fig.~\ref{Fig2}~b)]. By measuring the local magnetization one can observe that there is several harmonics participating in the signal. During the process of recrystallization, the resonances are suppressed and the system forms a DTC with period $2T$. From the perspective of a quantum walk, in this case the dynamics is restricted to two configurations for values of $\lambda$ close to one [see Fig.~\ref{Fig2}~c)]. 

Experimentally, one can initially prepare the system in a given configuration of spins polarized along the z-direction. Motivated by recent experimental realizations of DTCs with trapped ions, we consider the initial state $\ket{\psi(0)}=\ket{0,0,\dots,0}$. After $n$ periods of the drive, the density matrix of the system reads
$\hat{\rho}(nT)=\ket{\psi(nT)}\bra{\psi(nT)}$, where $\ket{\psi(nT)}=\sum_l A_l(nT)\ket{l}$. The dynamics of the populations $|A_l(nT)|^2$ allows one to access the information of the quantum walk in the configuration space. This geometrical representation is strongly related to localization properties of the Floquet states and their fractal dimension. When the fractal dimension is small, the system can only explore small regions of the configuration space, as we depict in Fig.~\ref{Fig7}~a). There one can see that the system explore just four configurations of the Hilbert space. By increasing the metamorphism parameter $\lambda$ we see that the time crystal melts and the system can visit more configurations as in Fig.~\ref{Fig7}~b). At the end of the recrystallization process for $\lambda\approx 1$ the system is only able to move between two configurations as it is illustrated in Fig.~\ref{Fig7}~c).

\section{Conclusions}

In this work we have shown that crystal metamorphism, a common phenomena in \textit{space} crystals, can also exist in the quantum realm of discrete \textit{time} crystals. In order to mimic and study this we have presented a system's Hamiltonian containing a deformation parameter, $\lambda$, that connects two time crystals of different order: a $2T$-DTC and a $4T$-DTC. We explicitly show how increasing values of $\lambda$ melts the time-crystal and makes its crystal order disappear. Thus, a system that was initially ordered and whose spectrum was very much localized, slowly transitions into an ergodic phase that now exhibits a high level repulsion. Crucially, as $\lambda$ approaches unity, the systems slowly starts to recrystallize and a phase transition occurs between the ergodic and the new time-crystalline phase.
Our model is not only capable of capturing such an interesting phenomena but it could also be experimentally realized with current technology. For this we have proposed an experimental protocol that exploits a quantum walk in the configuration space to allow for the measurement of the relevant quantities for the observation of quantum metamorphism. We believe NISQ devices~\cite{ippoliti2020many} to be an excellent test-bed to study crystalline phenomena in the dimension of time and we foresee our proposal as a promising candidate to explore fundamental quantum phenomena of such an exotic state of matter. 

\textit{Acknowledgement:\textemdash}
V.M.B acknowledges F. Wuman. We thank  A. Sakurai for valuable discussions. 
This work was supported in part from the Japanese MEXT Quantum Leap Flagship Program (MEXT Q-LEAP) Grant No. JPMXS0118069605, and the JSPS KAKENHI Grant No. 19H00662.

\bibliography{MBL,Mybib}

\end{document}